\def\RMP{{Rev. Mod. Phys.\ }\/}
\def\PRL{{ Phys. Rev. Lett.\ }\/}
\def\PRB{{ Phys. Rev. B\ }\/}

\def\etal{{\it et al.~}\/}

\def\be{\begin {equation}}
\def\ee{\end {equation}}
\def\ber{\begin {eqnarray}}
\def\eer{\end {eqnarray}}
\def\bers{\begin {eqnarray*}}
\def\eers{\end {eqnarray*}}

\makeatletter

\newcommand{\Rmnum}[1]{\expandafter\@slowromancap\romannumeral #1@}
\newcommand{\uvec}[1]{\boldsymbol{\hat{\textbf{#1}}}}
\makeatother

\makeatletter
\newcommand*\env@matrix[1][*\c@MaxMatrixCols c]{%
  \hskip -\arraycolsep
  \let\@ifnextchar\new@ifnextchar
  \array{#1}}
\makeatother

\documentclass[aps,prb,showpacs,superscriptaddress,a4paper,twocolumn]{revtex4-1}
\usepackage{amsmath}
\usepackage{float}
\usepackage{color}

\usepackage[normalem]{ulem}
\usepackage{soul}
\usepackage{amssymb}

\usepackage{amsfonts}
\usepackage{amssymb}
\usepackage{graphicx}
\begin {document}


\title{Intertwined non-trivial band topology and giant Rashba spin splitting} 

\author{Chiranjit Mondal}
\email{nilcm90@gmail.com}
\affiliation{Discipline of Metallurgy Engineering and Materials Science, IIT Indore, Simrol, Indore 453552, India}
\affiliation{Materials Modeling Group, Department of Physics, Indian Institute of Technology, Bombay, Powai, Mumbai 400076, India}

\author{Chanchal K. Barman}
\email{chanchalbarman91@gmail.com}
\affiliation{Materials Modeling Group, Department of Physics, Indian Institute of Technology, Bombay, Powai, Mumbai 400076, India}

\author{Aftab Alam}
\email{aftab@iitb.ac.in}
\affiliation{Materials Modeling Group, Department of Physics, Indian Institute of Technology, Bombay, Powai, Mumbai 400076, India}

\author{Biswarup Pathak}
\email{biswarup@iiti.ac.in }
\affiliation{Discipline of Metallurgy Engineering and Materials Science, IIT Indore, Simrol, Indore 453552, India}
\affiliation{Discipline of Chemistry, School of Basic Sciences, IIT Indore, Simrol, Indore 453552, India}

\date{\today}

\begin{abstract}

Composite quantum compounds (CQCs) have become an important avenue for the investigation of inter-correlation between two distinct phenomenon in physics. Topological superconductors, axion insulators etc. are few such CQCs which have recently drawn tremendous attention in the community. Topological nontriviality and Rashba spin physics are two different quantum phenomena but can be intertwined within a CQC platform. In this letter, we present a general symmetry based mechanism, supported by \textit{ab-initio} calculations to achieve intertwined giant Rashba splitting and topological non-trivial states simultaneously in a single crystalline system. Such co-existent properties can further be tuned to achieve other rich phenomenon. We have achieved Rashba splitting energy ($\Delta E$) and Rashba coefficient ($\alpha$) values as large as 161 meV and 4.87 eV$\AA$ respectively in conjunction with Weyl semimetal phase in KSnSb$_{0.625}$Bi$_{0.375}$. Interestingly, these values are even larger than the values reported for widely studied \textcolor{black}{topologically trivial} Rashba semiconductor BiTeI. The advantage of our present analysis is that one can achieve various topological phases without compromising the Rashba parameters, within this CQC platform. 

\end{abstract}


\maketitle


Simultaneous occurrence of two different quantum phenomena in a single material often provide the platform to understand the fundamental inter correlation between the two different processes in physics. The materials which have such combined quantum properties are called composite quantum compounds(CQCs).\cite{CQC} So far, composite phase between superconductivity and topology have been explored in the field of topological superconductivity.\cite{Hasan-Kane-RMP, Zhang-RMP} Topological axion insulator is another example of CQ phase between topology and magnetism, which is currently one of the emerging field of interest.\cite{axion-1, axion-2} 

Topological non-trivial band ordering and Rashba spin splitting could be yet another example of CQC. Though these two are distinct quantum phenomena, they can be intertwined through the common requirement of crystal symmetries and spin-orbit coupling (SOC). So far, topological insulator (TI) and ferroelectricity have been coupled in LiGaGe-type polar compounds for which strong SOC and broken inversion symmetry ($\mathcal{IS}$) are shown to be the two essential prerequisites.\cite{rashba_weyl_dirac} These two requirements also pave the path to realize another CQC between non-trivial topology (in the presence of strong SOC) and giant Rashba spin splitting (which requires both inversion breaking and strong SOC).Though the reasonably large Rashba splitting and TI phase have already been found in polar semiconductor BiTeI, it requires external pressure and fine tuning of electronic structure.\cite{nat-mat-BiTeI, TI-BiTeI-1, TI-BiTeI-2} Nevertheless, a general mechanism or criterion to realize such CQ phase between topological non-trivial band order and Rashba spin physics has never been explored. Here, we report a definite pathway to achieve such CQ phase involving Rashba spin physics and topological non-triviality. In particular, we disclose a material platform that can possess non-trivial Weyl semimetal (WSM) and TI phase along with large Rashba splitting.

In this letter, we report giant Rashba co-efficient ($\alpha$) value in a polar compound KSnSb$_{1-x}$Bi$_x$ which simultaneously possess either WSM or TI phase depending on the value of $x$. Using a combined study involving group theoretical analysis and \emph{ab-initio} calculation (see Sec. I of supplemental material (SM)\cite{supp} for computational details), we explained the origin of such giant Rashba splitting. Our discussions enlighten a general and necessary guidelines to achieve such a composite phase between Rashba splittiing and topological non-triviality.

\begin{figure}[t!]
\centering
\includegraphics[width=\linewidth]{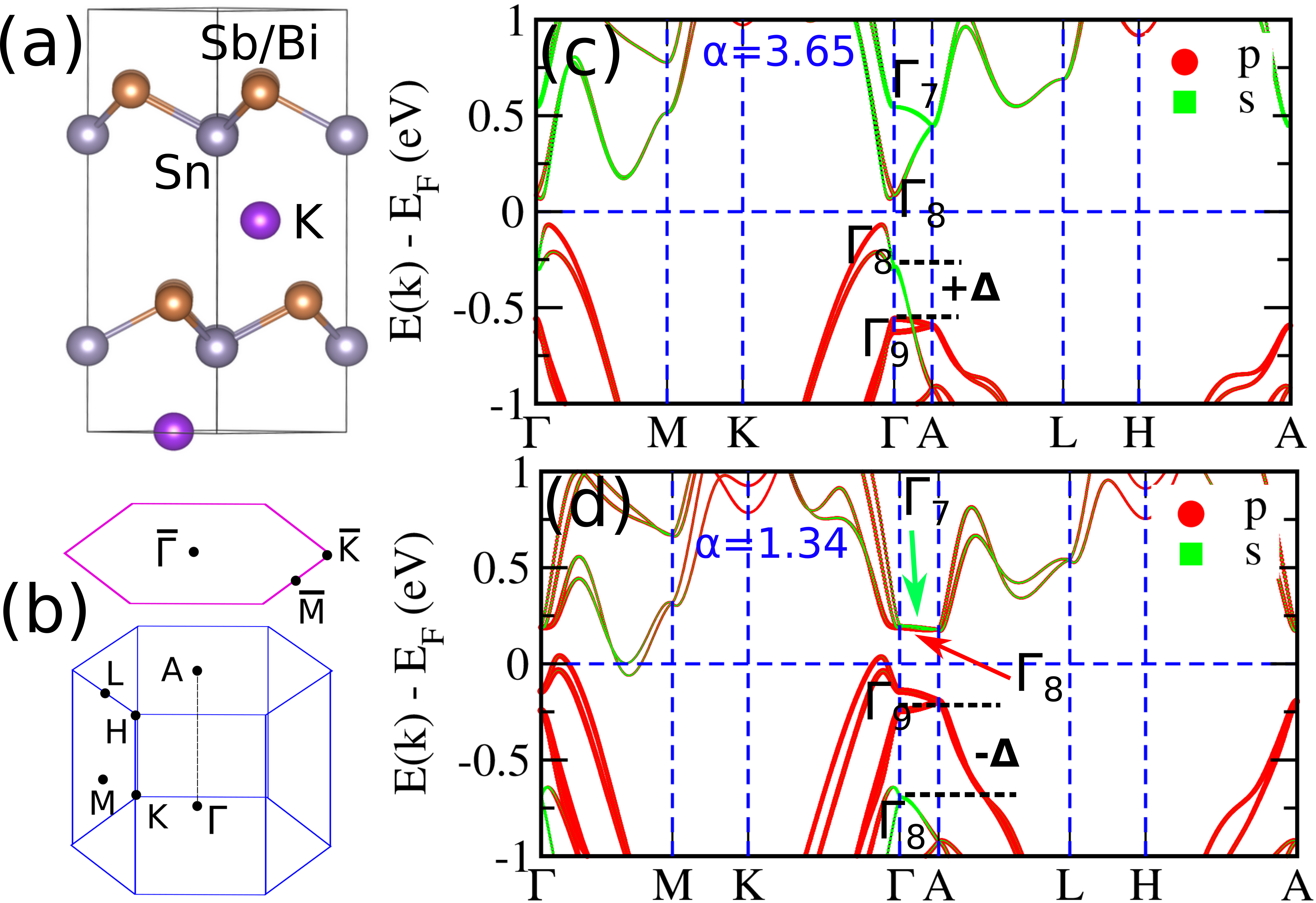}
\caption{(Color online) (a) Crystal structure of KSnSb and KSnBi. (b) Bulk and surface Brillouin zones. Orbital projected PBE+SOC bulk band structure of (c) KSnSb and (d) KSnBi.  $\alpha$ is the value of Rashba coefficient in eV$\AA$ unit. $\Gamma_i$'s indicate the band character at $\Gamma$ point in the BZ.}
\label{fig1}
\end{figure}



The polar compound KSnSb (KSnBi) belongs to space group P6$_3$mc which primarily consist of a $sp_3$ type buckled honeycomb layer with [SnSb]$^{-\delta}$ ([SnBi]$^{-\delta}$) charge configuration. [K]$^{+\delta}$ (where, $\delta<1$) serve as stuffing layers and generate the net polarization along z-axis, as shown in Fig.~\ref{fig1}(a). Further details of charge states on various atoms and the associated imbalance are discussed in SM.\cite{supp}
The optimized lattice parameters for KSnSb (KSnBi) are $a=4.45 ~\AA$, $c=13.28 ~\AA$ ($a=4.56 ~\AA$, $c=13.34 ~\AA$) which are in close agreement with experiments.\cite{KSnSbExpt} The dynamical stability of both these compounds are also confirmed via the absence of imaginary phonon frequencies.\cite{phonon} The oppositely charged alternate layers in the crystal structure create a local potential imbalance. This, in turn, results in momentum dependent large Rashba type spin splitting in M-$\Gamma$-K plane of the Brillouin zone (BZ) (see Fig.~\ref{fig1}(b)) in both the compounds. Figure~\ref{fig1}(c,d) shows the band structure of KSnSb and KSnBi with SOC using the Perdew-Burke-Ernzerhof (PBE) functional. The value of $\alpha$ in valence band (VB) for KSnSb (KSnBi) is 3.65(1.34)eV$\AA$ (see Ref.~[\onlinecite{RASHBA}] for pictorial representation of Rashba splitting and Ref. [\onlinecite{deltaE-deltaK}] for detailed calculation of $\alpha$ in terms of Rashba splitting energy $\Delta E$).  This goes against the natural expectation; as Bi has much larger SOC than Sb and hence expected to have larger $\alpha$ value in KSnBi. We will resume and explain the origin of such outcome in the subsequent sections, and also provide a general guidance to achieve high Rashba splitting using symmetry adopted analysis.

\begin{figure}[t]
\centering
\includegraphics[width=0.9\linewidth]{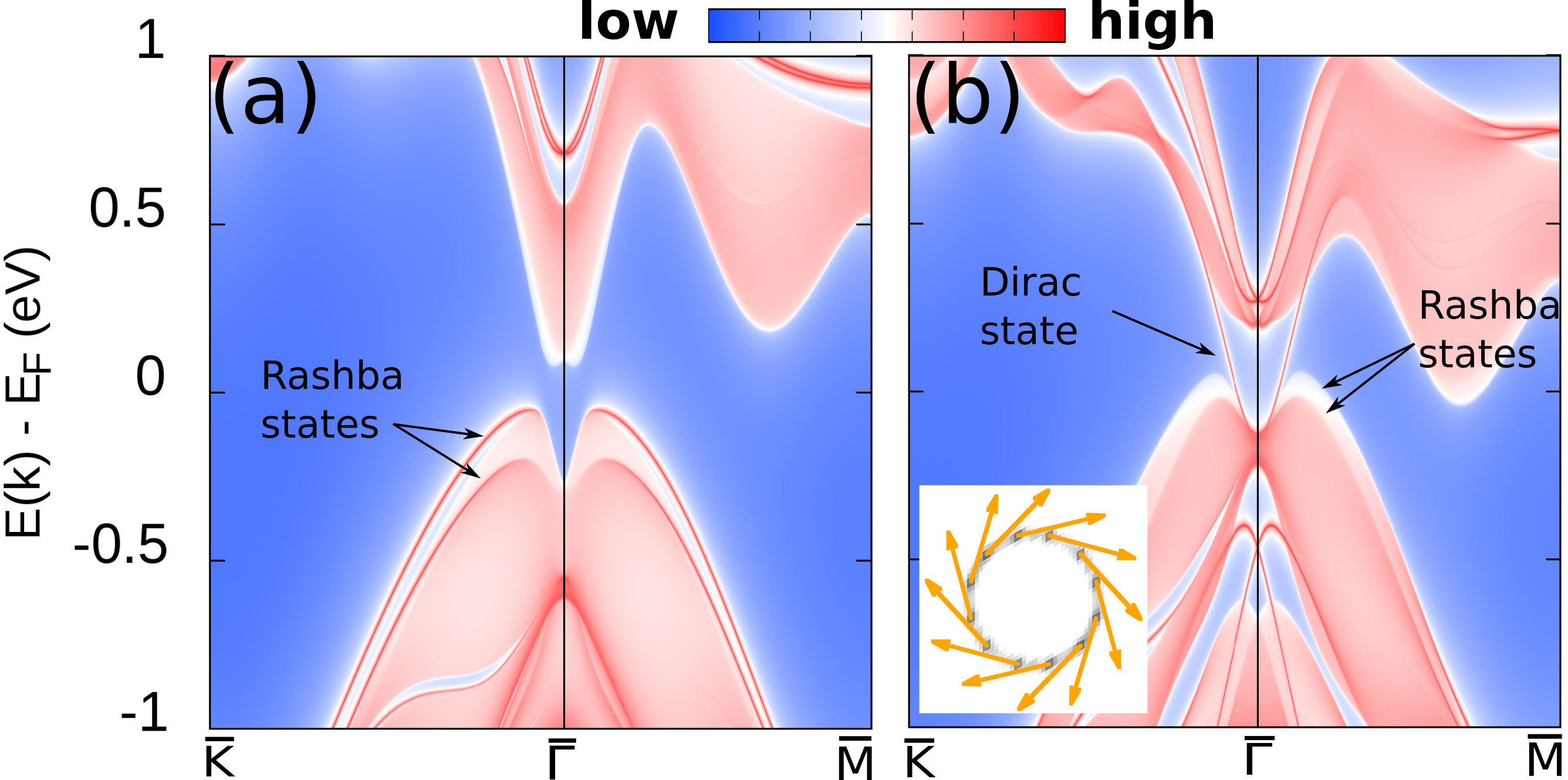}
\caption{(Color online) (001) projected surface states (SSs) of (a) KSnSb and (b) KSnBi. Inset in (b) shows the topological Dirac Fermi arc (FA) with helical spin texture for KSnBi.}
\label{fig2}
\end{figure}

In Fig.~\ref{fig1}(c,d), the nature of bands at $\Gamma$ point are labeled according to the irreducible representations (IRs) of $C_{6v}$ point-group symmetry (see Table S2-S4 in SM for point group character tables). The conduction band minima (CBM) for both the compounds are dominated by $s$ like $\Gamma_8$ character. The representations of first  and second valence band maxima (VBM) of KSnSb  are $s$-like $\Gamma_8$ and $p$-like $\Gamma_9$ respectively. The gap between the two maxima ($\Gamma_8$ and $\Gamma_9$) is the band inversion strength (BIS), denoted by $\pm\Delta$. For KSnSb, the BIS is positive and hence it is a trivial insulator, while KSnBi is a TI (with $-\Delta$), see Fig.~\ref{fig1}(d). Further detail on symmetry labeling of the bands and band inversion mechanism have been discussed in SM section III. We have also validate the accuracy of the PBE results using highly accurate HSE06 functional (see Fig.S1 of SM.\cite{supp}). Apart from small rearrangements, the band ordering in the two compounds remain the same as that of PBE+SOC prediction. Hence, here after, we have considered only PBE functional for further calculations. We have simulated the $\mathbb{Z}_2$ index and surface states (SSs) for both the  compounds. The spectral flow of Wannier charge centers on $k_z =0$ plane reconfirms the topological band inversion at the $\Gamma$ point for KSnBi, whereas KSnSb has trivial $\mathbb{Z}_2$ index. These results are shown in Sec. III of SM.\cite{supp} Non-trivial nature of KSnBi is further confirmed by the presence of Dirac SSs with helical spin texture, see Fig.~\ref{fig2}(b). KSnSb, however, show trivial SSs (Fig.~\ref{fig2}(a)).

We shall now discuss the nature of band degeneracy and Rashba spin splitting from the context of structural symmetry and polar axis concept. The point group symmetry at $\Gamma$-point and along $\Gamma$-A direction is $C_{6v}$. So, the simultaneous presence of $C_6$, $C_2$, 3 $\sigma_v$ and 3 $\sigma_d$ enforce all the states to be at least doubly degenerate along $\Gamma$-A. States at all other points, except eight time reversal invariant momentum (TRIM) points, are singly degenerate. The polar z-axis ensure maximum band splitting in the plane which is perpendicular to the polar axis. The value of momentum offset ($\Delta$k) and splitting energy ($\Delta$E) for KSnSb are 0.078 $\AA^{-1}$ and 142 meV along $\Gamma$-K direction, while those for KSnBi are 74 meV and  0.110 $\AA^{-1}$ respectively. Now, let us resurrect our discussion about  getting an unexpectedly large value of $\alpha$ in KSnSb compared to KSnBi, despite having a larger SOC strength in the later. To elucidate this, we write a general \textit{k.p} Hamiltonian using perturbation theory as (around $\Gamma$ point); 

\begin{align*}
H({\bf k}) =\;& H(0)+\frac{\hbar^2 k^2}{2m_0}+\frac{\hbar}{m_0}{\bf k\cdot p}+\frac{\hbar}{4 m_0^2 c^2}[{\bf \nabla{V}\times p}]\cdot\sigma + \\ & \frac{\hbar^2}{4 m_0^2 c^2}[{\bf \nabla{V}\times k}]\cdot\sigma,
\end{align*}

where {\bf V, p} and $\sigma$ represent the crystal potential, momentum operator and Pauli matrix respectively. The 4th and 5th terms in the Hamiltonian are SOC terms which have been added as a perturbation that operate on the Bloch periodic function. The last {\bf k}-dependent term is analogous to the motion of wave packet. Given the fact that the velocity of wave packet is much smaller than the velocity of electrons in atomic orbital, we can neglect the last term. The coupling between 3rd and 4th terms provide the second order energy correction\cite{C6v-Ham} as;\\ {\small $\Delta E_n^{(2)}\propto \sum_{m\neq n}\dfrac{\langle u_n|(\frac{\hbar}{4 m_0^2 c^2}[{\bf \nabla{V}\times p}]\cdot\sigma)|u_m\rangle \langle u_m|{\bf k\cdot p}|u_n\rangle +c.c}{E_n - E_m}$}     This energy correction term clearly indicates that the spin splitting does not only depend on the strength of SOC but also on two other important factors, (i) energy difference ($E_n-E_m$) between two bands and (ii) the character of the Bloch states corresponding to two bands i.e. spin splitting is possible only if the term $\langle u_n|\frac{\hbar}{4 m_0^2 c^2}[{\bf \nabla{V}\times p}]\cdot\sigma|u_m \rangle$ is symmetrically allowed. As such, the compounds having large SOC strength will produce giant spin splitting in two closely spaced bands only if they have symmetrically same characters. Now, coming back to the crystal symmetry of KSnSb (KSnBi), we can readily construct the Rashba type Hamiltonian imposing the symmetry invariance on the {\bf k.p} Hamiltonian. Around $\Gamma$ point, the representation of $C_n$($n$=2, 3 \& 6), $\sigma_v$ and $\sigma_d$ have $\mathrm{e}^{-i\sigma_z\pi/n}$, $i\sigma_x$ and $i\sigma_y$ symmetry respectively, where $\sigma_{x,y,z}$ are the Pauli matrices for spin degree of freedom (DOF). Using the time reversal symmetry ($\mathcal{TRS}$) operator $i\sigma_y K$ for spin-1/2 particle, one can find the linear order Rashba Hamiltonian as;\cite{C6v-Ham} 
\begin{eqnarray}\label{H_c6v}
H_R^{C_{6v}}=\alpha(k_x\sigma_y-k_y\sigma_x) 
\end{eqnarray}

\begin{figure}[t]
\centering
\includegraphics[width=\linewidth]{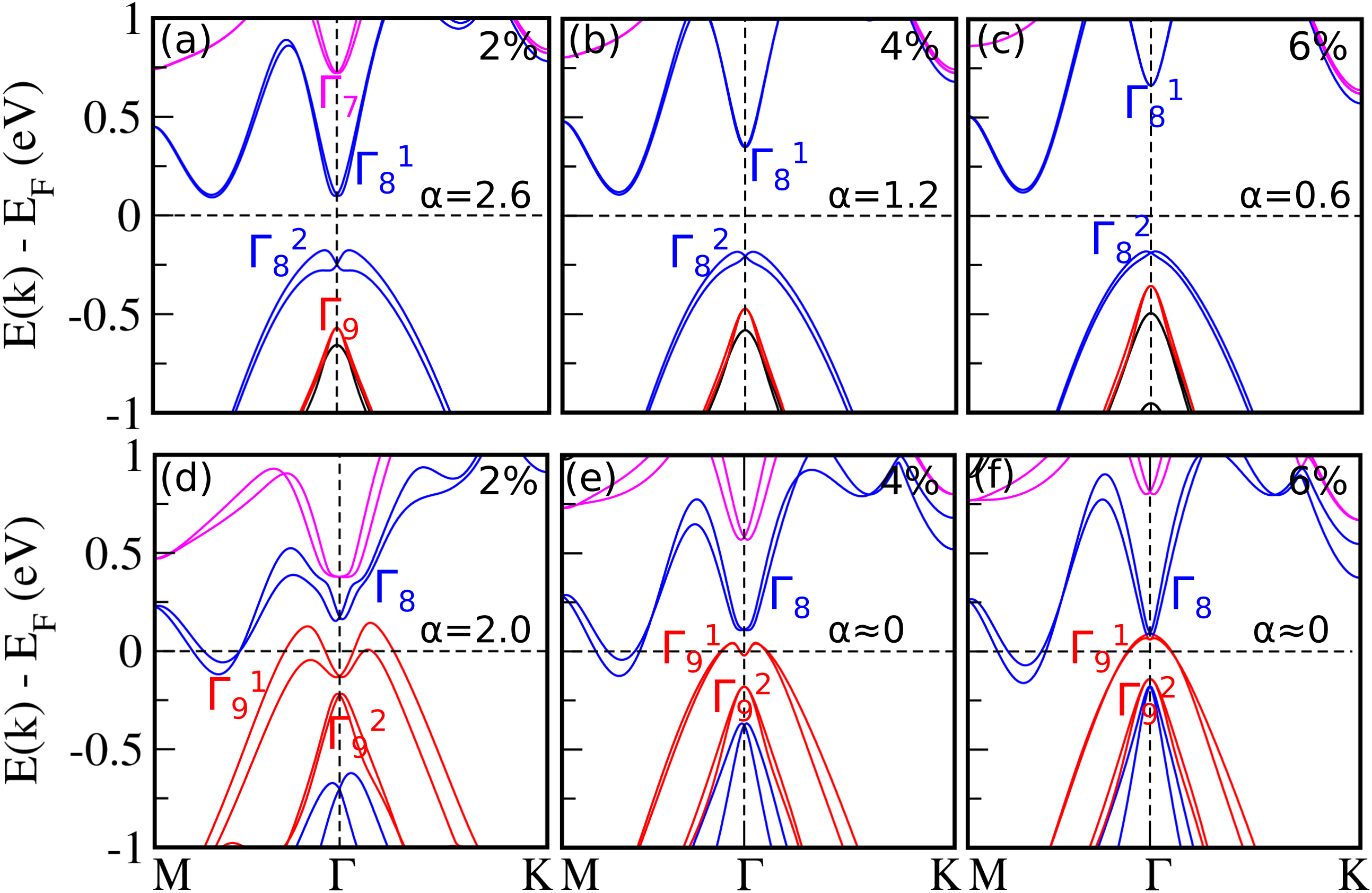}
\caption{(Color online)Bulk band structures of (a-c) KSnSb and (d-f) KSnBi under 2\%, 4\% \& 6\% compressive hydrostatic pressure (HP).  $\alpha$ is the value of Rashba coefficient in eV$\AA$. $\Gamma^{n}_i$'s indicate the irreducible representations of bands at $\Gamma$ point. The prefix \emph{n}(=1,2) represents band index.}
\label{fig3}
\end{figure}

{\par} Now, if we closely look at the band structure of KSnSb and KSnBi (Fig.~\ref{fig1}), the CBM and VBM of the former has the same character ($\Gamma_8$) whereas the later has $\Gamma_8$ character for CBM and $\Gamma_9$ for VBM. As such, the term $\langle u_n|\frac{\hbar}{4 m_0^2 c^2}[{\bf \nabla{V}\times p}]\cdot\sigma|u_m \rangle$ is symmetrically allowed for KSnSb at  $\Gamma$ point , but not for KSnBi. Therefore, even though KSnBi has larger SOC strength, KSnSb provides larger Rashba splitting because of the dominance of symmetry related term.

To further demonstrate the influence of energy difference ($E_n-E_m$) between two bands of different band index and symmetry of those two bands on the Rashba energy $\Delta E$ and Rashba coefficient $\alpha$, we calculate the band structures of both KSnSb and KSnBi under external pressure. We apply hydrostatic pressure (HP) (which maintains the original crystal symmetry $C_{6v}$) by reducing the lattice constants by 2$\%$, 4$\%$ and 6$\%$. The corresponding band structures are shown in  Figs.~\ref{fig3}(a-c) for KSnSb and \ref{fig3}(d-f) for KSnBi respectively. The bands are denoted by the IRs of $C_{6v}$ at the $\Gamma$ point. It is to be noted that KSnBi still produce a reasonably large splitting in VBM ($\alpha=2.0$) because of the presence of another $\Gamma_9$ band ($\Gamma_9^2$) just below the VBM ($\Gamma_9^1$). With increasing pressure, the band gap (E$_g$) of KSnSb increases (Fig.~\ref{fig3}(a-c)). However, the band gap (at $\Gamma$ point) for KSnBi decreases with increasing pressure (Fig.~\ref{fig3}(d-f)). This is due to the inverted band order i.e, the ``$-$ve'' band gap of KSnBi; and compressive pressure gradually decreases the band gap which leads to a trivial insulating band order. With increasing pressure in KSnBi, the energy difference (i.e, $E_n-E_m$) between two $\Gamma_{9}$ ($\Gamma_{9}^{1}$ and $\Gamma_{9}^{2}$) valence bands increases. As a result, the Rashba energy ($\Delta E$) decreases  giving rise to a decrease in $\alpha$ value, as evident from Fig.~\ref{fig3}(d-f). It is extremely important to note that there is no direct correlation between $\alpha$ and the band gap (E$_g$). Rather $\alpha$-value  is purely dictated by the splitting energy ($\Delta E$) which, in turn, depends on the energy difference ($E_n-E_m$) between two bands of same character.\cite{deltaE-deltaK} See section IV of SM \cite{supp} for further discussion on the evolution of energy separation under pressure, and hence the $\alpha$-value.

\begin{figure}[t]
\centering
\includegraphics[width=\linewidth]{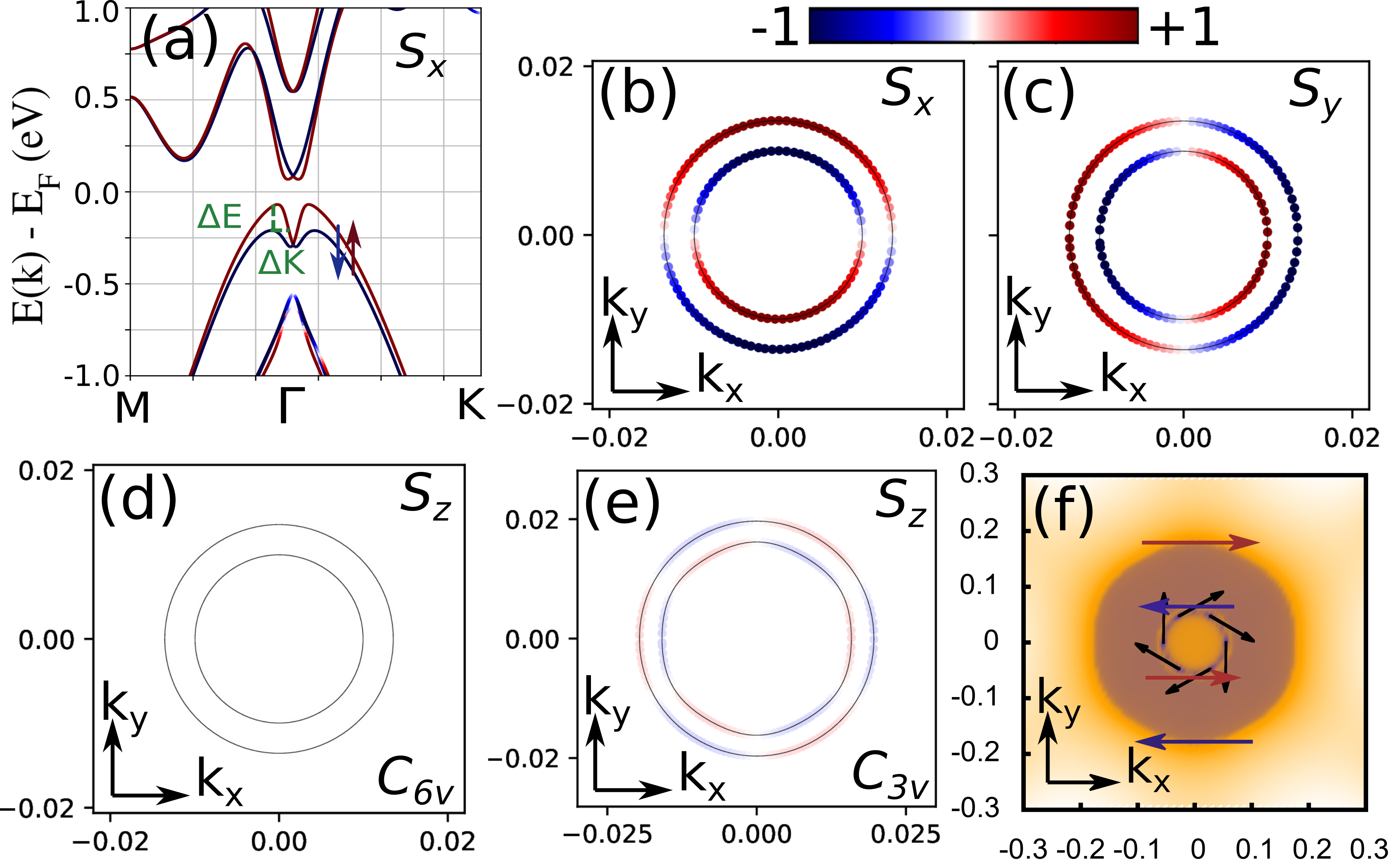}
\caption{(Color online) Rashba spin texture of KSnSb and KSnBi. (a) Projected in-plane $S_x$ spin components on the band structures of KSnSb along K-$\Gamma$-M direction. (b-d) Constant energy contour at 250 meV of the in-plane $S_x$, $S_y$ and out of plane $S_z$ components for KSnSb. (e) $S_z$ component with hexagonally distorted contour for KSnSB$_{0.5}$Bi$_{0.5}$, caused by the reduction of symmetry from $C_{6v}$ to $C_{3v}$. (f) Topological surface spin texture of KSnBi (shown by black arrows in the inner circle) and schematic representation of Rashba spin by red and blue arrows on the projected bulk states.}
\label{fig4}
\end{figure}

Next, we extend our discussion to the Rashba spin polarization as the hallmark signature of Rashba bands. Figure~\ref{fig4}(a) shows the projection of $S_x$ spin components on the band  structure of KSnSb (at ambient condition) along K-$\Gamma$-M direction. As expected, both the inner and outer branches are completely spin polarized and the spin helicity have same sign for both CB and VB, i.e, the inner (outer) branch of both CB and VB have low (high) spin value. This feature is in complete contrast to the extensively studied Rashba semiconductor BiTeI, where spin textures have opposite chirality in VB and CB. \cite{BiTeI-spintex} Figure~\ref{fig4}(b-d) shows projected $S_x$, $S_y$ and $S_z$ spin components on the constant energy contours (CECs) at 250 meV, clearly indicating the purely in-plane nature of the net spin polarization. The out-of-plane $S_z$ component is zero and both the inner and outer branch of CECs are perfectly circular. This can be explained from Rashba Hamiltonian (Eq.~\ref{H_c6v}) that allows only linear order terms in \textbf{k} and does not involve $\sigma_z$ spin matrix. Nevertheless, a finite out-of-plane component can be induced over the CECs by reducing the structural symmetry from C$_{6v}$ to C$_{3v}$. This is because the Rashba Hamiltonian under C$_{3v}$ allows non-zero cubic terms involving $\sigma_z$ matrix. The Hamiltonian in this case is given by,\cite{TI-BiTeI-1} 
\begin{eqnarray}\label{H_c3v}
H_R^{C_{3v}}=\alpha(k_x\sigma_y-k_y\sigma_x) + \lambda (3k_x^2 - k_y^2)k_y \sigma_z.
\end{eqnarray}
The cubic terms not only allow the finite out-of-plane component but also provide a hexagonal distortion in CEC. To explicitly show this, we have computed the Rashba spin texture of KSnSb$_{0.5}$Bi$_{0.5}$ which possess C$_{3v}$ symmetry. Figure~\ref{fig4}(e) shows the S$_{z}$ component of KSnSb$_{0.5}$Bi$_{0.5}$ at 250 meV in which the inner branch  indeed shows a hexagonal distortion. This distortion arises due to the trigonal wrapping of S$_{z}$ component associated with the cubic term, as clearly visible in Fig.~\ref{fig4}(e). To understand this from the structural chemistry point of view, we write the basic Rashba Hamiltonian as; $H_R$=$\frac{\hbar}{4 m_0^2 c^2}[{\bf \nabla{V}\times p}]\cdot\sigma$. Now, let us consider the perfectly two dimensional free electron gas layers that form in the crystal structure of KSnSb (KSnBi) by the alternate layers of positive [K]$^{+2}$ and negative [SnSb]$^{-2}$ ([SnBi]$^{-2}$) charge clouds such that, the direction of potential gradient is strictly oriented along the z-axis and there is no in-plane anisotropy of charges. For such a situation, the Rashba Hamiltonian takes the form (in x-y plane); $H_R$=$\alpha[{\bf \uvec{k}\times p_\parallel}]\cdot\sigma$, where $\bf \uvec{k}$ is the unit vector along potential gradient (z-axis) and $\bf p_\parallel$ is the in-plane wave vector. This simple model nicely reproduces our $\textit{ab-initio}$ result of 100$\%$ spin polarization of in-plane ($S_x$,$S_y$) components and zero spin polarization of  out-of-plane component. The in-plane components are axially symmetric and the signature of $\mathcal{TRS}$, (i.e, $\bf S(p_\parallel)$ = $\bf -S(-p_\parallel)$ and $\bf E(p_\parallel)$ = $\bf E(-p_\parallel)$) are prominent in the spin textures, see Fig.~\ref{fig4}(b-c). In contrast, KSnSb$_{0.5}$Bi$_{0.5}$  (with C$_{3v}$ symmetry) possibly acquires an in-plane potential gradient due to its relatively lower symmetry. This in-plane anisotropy induces an out-of-plane S$_{z}$ component on CECs, see Fig.~\ref{fig4}(e). The in-plane components of KSnSb$_{0.5}$Bi$_{0.5}$ are shown in  SM.\cite{supp} Figure~\ref{fig4}(f) shows the projected surface and bulk bands on (001) plane of KSnBi. As KSnBi is an TI, the Fermi surface (FS) contains both the topological spin (inner circle with black arrows) and the Rashba spin over the bulk CECs (outer dark bulk states with red and blue arrows). The FS is simulated just below the VBM to get both these features.

\begin{figure}[t]
\centering
\includegraphics[width=\linewidth]{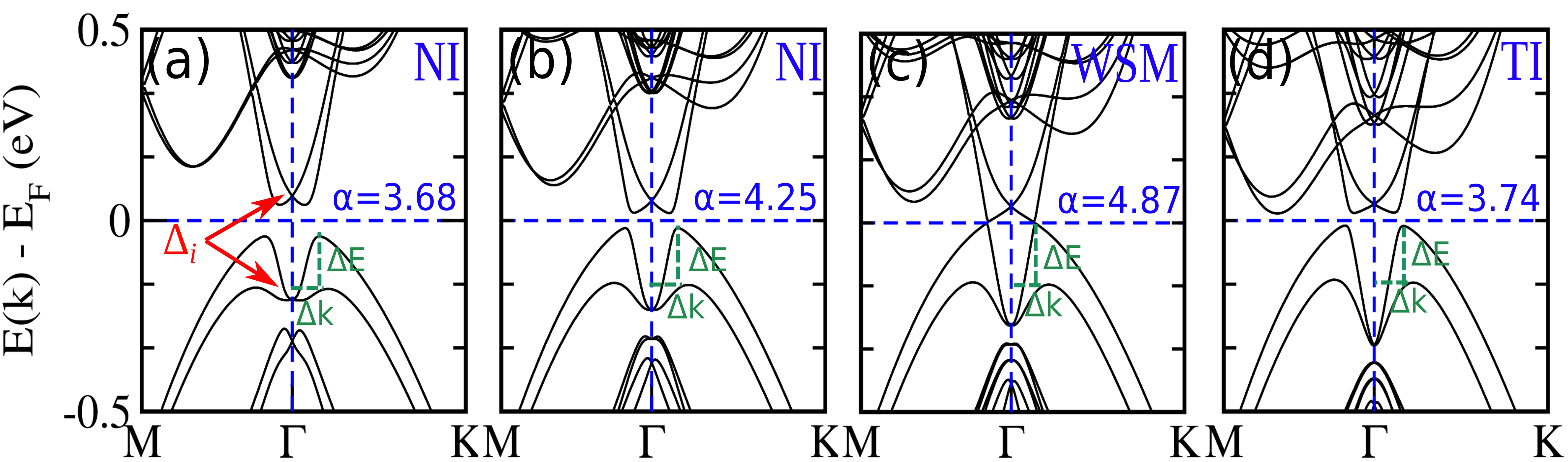}
\caption{(Color online)  Bulk band structures of KSnSb$_{1-x}$Bi$_x$ for (a) $x=0.125$, (b) $x=0.25$, (c) $x=0.375$ and (d) $x=0.5$. NI, WSM and TI stands for topologically trivial, Weyl semimetal and topological insulator phases respectively. $\Delta_i$ represents the IRs of bands at $\Gamma$ point.}
\label{fig5}
\end{figure}

\begin{table}[b]
\begin{ruledtabular}
\caption{Rashba splitting energy ($\Delta E$), momentum offset($\Delta K$), Rashba coefficient($\alpha$) and formation energy($\Delta$E$_f$)\cite{chemicalstability} of different materials. NI, WSM and TI indicates topologically trivial, Weyl semimetal, and topological insulator phases respectively.}
\label{Table1}
\begin{tabular}{c c c c c c c c c c}
& Compound & $\Delta E$ & $\Delta k$     & $\alpha$ & T-phase  & \textcolor{black}{$\Delta$E$_f$}      \\ 
&  & (meV)  & ($\AA$)      & (eV$\AA$) &  & \textcolor{black}{(meV/atom)}           \\ \hline
& KSnSb$^a$  &    142   &  0.078  &      3.65         & NI  & \textcolor{black}{-356.9} \\
& KSnSb$_{0.875}$Bi$_{0.125}$$^a$                       &    138   &  0.075  &      3.68         & NI  & \textcolor{black}{-347.3}  \\
& KSnSb$_{0.75}$Bi$_{0.25}$$^a$                         &    149   &  0.070  &      4.25         & NI   & \textcolor{black}{-338.0}  \\
& KSnSb$_{0.625}$Bi$_{0.375}$$^a$                       &    161   &  0.066  &      4.87         & WSM   & \textcolor{black}{-328.9}  \\
& KSnSb$_{0.5}$Bi$_{0.5}$$^a$                          &    148   &  0.079  &      3.74         & TI    & \textcolor{black}{-320.1}  \\
& KSnBi$^a$                                            &    74    &  0.110  &      1.34         & TI    & \textcolor{black}{-291.2}  \\ 
& BiTeI$^b$                                            &    100   & 0.052   &      3.8          & NI     &  \\
\end{tabular}
\end{ruledtabular}
$^a$[This Work], $^b$[Ref. \onlinecite{nat-mat-BiTeI}]
\end{table}

We now ask an important question; how to engineer the band structure so as to retain its topological non-trivial behavior, but enhance Rashba splitting. Obviously, external pressure can tune the band topology, as seen in Fig.~\ref{fig3}, but because of the symmetry constraint, the band splitting (as well as $\alpha$) reduces significantly. To bypass the symmetry constraints on Rashba splitting, we adopt the alloying induced symmetry lowering mechanism. Lower symmetric structures allow less number of IRs for defining band characters and hence the CBM and VBM are more probable to have same band characters, that in turn allows higher Rashba splitting. We simulated mixed compounds KSnSb$_{1-x}$Bi$_x$ ($x$=0.125,0.25,0.375 and 0.5) using a $2\times2\times2$ supercell of a 6-atom unit cell, whose band structures are sown in Fig.~\ref{fig5} (a-d). For all these $x$ values, the structural symmetry reduces to $C_s$ from $C_{6v}$. In contrast to $C_{6v}$, $C_s$ only has two IRs $\Gamma_1$ and $\Gamma_2$ (say). These two IRs will form Kramer's degeneracy because of $\mathcal{TRS}$ at all the TRIM points. Hence, all the bands of KSnSb$_{1-x}$Bi$_x$ have same character $\Delta_i$ at $\Gamma$ point which removes the symmetry restrictions of the Rashba band splitting and hence increase the value of $\alpha$. With increasing $x$ (Bi concentration), SOC strength increases and lattice expands as Bi is bigger than Sb. This, in turn, decreases the band gap. The gap closes at $x=0.375$ and reopen again through a topological phase transition, see Fig.~\ref{fig5}. For $x=0.375$, the system becomes gapless and host the novel topological WSM phase. Beyond that, at $x=0.5$, a topological gap reopens and system becomes a strong TI. Notably, the value of $\alpha$ increase as $x$ increases until $x=0.375$ where the system becomes a gapless WSM, achieving the largest value of $\alpha$(=4.87 eV$\AA$). Beyond this concentration ($x$), $\alpha$ decreases. This trend of $\alpha$ arises due to the different nature of band evolution, quantifying $\Delta E$, above/below the topological phase transition point. Even for TI phase ($x=0.5$), the $\alpha$ value is reasonably high (3.74 eV$\AA$). In Table~\ref{Table1}, we listed all the Rashba parameters in VB for different values of $x$ and compared them with the previously studied Rashba material BiTeI. A similar data for CB have been presented in SM.\cite{supp} Formation energies ($\Delta$E$_f$)\cite{chemicalstability} and the nature of topological phase of  these compounds are also listed. We have also validated our symmetry based arguments for high $\alpha$-value of BiTeI\cite{nat-mat-BiTeI, BiTeI-spintex, TI-BiTeI-1, BiTeI-PRL, BiTeI-ori} in  SM.\cite{supp}

In conclusion, we use first-principle simulation combined with group theoretical analysis to disclose the true origin of giant Rashba splitting in a crystalline system. Taking KSnSb$_{1-x}$Bi$_{x}$ as an example system, we show that the crystalline symmetry and the Rashba splitting energy between the bands play crucial role to maximize Rashba coefficient in a broken $\mathcal{IS}$ and strong spin-orbit coupled  system. We show that the crystals with lower symmetry are more favorable to yield larger Rashba splitting. We provide a simple yet viable scheme to simultaneously realize the co-existence of Rashba and topologically non-trivial properties in a single material. We achieved a record high value of Rashba coefficient ($\alpha=4.87$ $eV\AA$) in KSnSb$_{0.625}$Bi$_{0.375}$ which simultaneously possess non-trivial WSM phase. We are able to tune the topological phase to TI without much compromising the high Rashba parameter. Our proposed mechanism is robust in a sense that it does not require fine tuning of the system itself, but expected to be applicable to all class of compounds. We strongly believe that our findings in this letter are quite insightful along the line of Rashba and topological physics and will surely attract the experimentalists for future studies.

{\par} This work is financially supported by DST SERB (EMR/2015/002057), India. We thank IIT Indore for the lab and computing facilities. CM and CKB acknowledge MHRD-India for financial support. CKB thanks IIT Bombay spacetime computing facilities. 






\begin{thebibliography}{113}
\bibitem{deltaE-deltaK} Rashba splitting energy ($\Delta$E) and momentum offset ($\Delta$K) are the parameters which quantify the strength of Rashba parameter or Rashba coefficient ($\alpha$). Inclusion of SOC  decouple the spin bands, except at $\Gamma$ point where it is still degenerate due to the time reversal symmetry. The splitting energy and momentum offset, $\Delta$E and $\Delta$K, can be quantified from the band structure, as shown in Fig.~4 (a) \& Fig.~5 (a-d). The value of $\alpha$ can then be calculated using the formula; $\alpha$ = 2$\Delta$E/$\Delta$K.
\bibitem{chemicalstability} Formation energy is defined as, $\Delta $E$_f =E_{Comp} - \sum x_{i} E_{i} $, where $E_{Comp}$ is the total energy of the compound and $ E_{i} $ is the energy of the constituent elements in their equilibrium phase, all at their optimized lattice constants. $x_i$ is the proportion of $i^{th}$ element in the compound. $\Delta$E$_f $ is a measure of the chemical stability of the concerned compound.
\end{thebibliography}
\end{document}



\title{%
Supplemental Material for `` \large Intertwined non-trivial band topology and giant Rashba spin splitting"}

\author{Chiranjit Mondal}
\email{nilcm90@gmail.com}
\affiliation{Discipline of Metallurgy Engineering and Materials Science, IIT Indore, Simrol, Indore 453552, India}

\affiliation{Materials Modeling Group, Department of Physics, Indian Institute of Technology, Bombay, Powai, Mumbai 400076, India}

\author{Chanchal K. Barman}
\email{chanchalbarman91@gmail.com}
\affiliation{Materials Modeling Group, Department of Physics, Indian Institute of Technology, Bombay, Powai, Mumbai 400076, India}

\author{Aftab Alam}
\email{aftab@iitb.ac.in}
\affiliation{Materials Modeling Group, Department of Physics, Indian Institute of Technology, Bombay, Powai, Mumbai 400076, India}

\author{Biswarup Pathak}
\email{biswarup@iiti.ac.in }
\affiliation{Discipline of Metallurgy Engineering and Materials Science, IIT Indore, Simrol, Indore 453552, India}
\affiliation{Discipline of Chemistry, School of Basic Sciences, IIT Indore, Simrol, Indore 453552, India}

\date{\today}

\maketitle 


\beginsupplement

Here, we provide the details of computation in section \ref{computation}. We have discussed the charge state of KSnSb in section \ref{charge}. Band inversion mechanism have been discussed in section \ref{band-inversion}. HSE06 band structures are given in Sec.~\ref{HSE06}.
Calculation of $\mathbb{Z}_2$ index of KSnSb and KSnBi are shown in Sec.~\ref{Z2} to confirm their topological states. Band gap evolution under pressure for KSnSb and KSnBi are discussed in Sec.~\ref{band-gap-evoluation}. The in-plane components of spin texture for KSnSb$_{0.5}$Bi$_{0.5}$ are shown in section \ref{in-plane}. Finally, we have verified our proposed mechanism for getting large Rashba coefficient for BiTeI in section \ref{org}. In Table \ref{Table1}, we have listed the values of Rashba parameters for different compounds in conduction bands and their topological states. In section \ref{CT}, the character tables are shown.

\subsection{Computational Details}
\label{computation}
First principle calculations were carried out using Density Functional Theory (DFT) implemented within the Vienna Ab Initio Simulation Package (VASP)\cite{Hafner1993,Joubert1999}. Plane wave basis set using projector augmented wave (PAW)\cite{PEBLOCH1994} method was used with an energy cutoff of 500 eV. Generalized-gradient approximation by Perdew-Burke-Ernzerhof (PBE)\cite{Joubert1999} was employed to describe the exchange and correlation effect. Brillouin zone (BZ) integration was performed using a $12\times 12\times 4$ k-mesh. Total energy (force) was converged set upto 10$^{-6}$ eV (0.01 eV/\AA). To accurately probe the band order and the band gap, we also employed HSE06\cite{HSE} hybrid functional. The relativistic spin orbit coupling (SOC) was included in all the calculations. Maximally localized Wannier functions (MLWFs)\cite{mlwf1,mlwf2,mlwf3} obtained from wannier90 package\cite{w90}, were used to construct a tight-binding (TB) Hamiltonian. The topological properties including surface spectrum and Fermi arcs were calculated using the iterative Green's function\cite{greenfn1,greenfn2,greenfn3} approach implemented within Wannier-Tools package\cite{WTools}.

\begin{figure}[b!]
	\centering
	\includegraphics[width=1\textwidth]{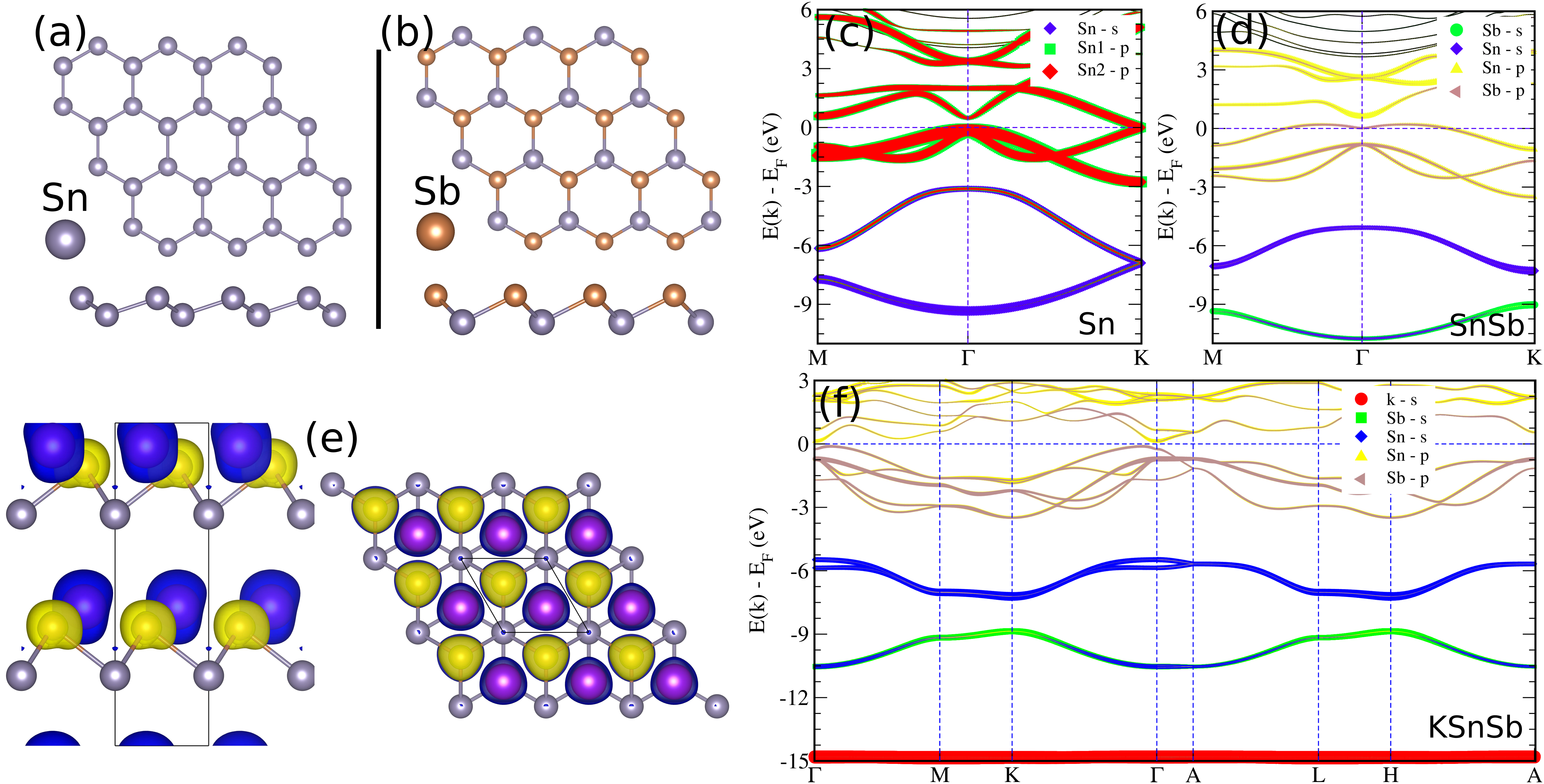}
	\caption{(Color online) Top and side view of (a) Sn and (b) SnSb monolayers. Band structures of (c) Sn, (d) SnSb without SOC. (e) Side and top views of relative charge density plot of KSnSb. (f) Band structure of KSnSb without SOC. The orbital projections are shown by different colors}
	\label{charge}
\end{figure}

\subsection{Charge state analysis}
\label{charge}
The electronic configurations of the constituent elements of these systems are: K: 3p$^6$ 4s$^1$ ; Sn: 4d$^{10}$ 5s$^2$ 5p$^2$ ; and Sb : 5s$^2$ 5p$^3$. Sn and Sb s-orbitals contribute at much lower energy, at/around -6 eV. They do not hybridized with other orbitals which indicate that the s-orbitals of Sn and Sb do not contribute to the chemical bonding. However, Sn and Sb p orbitals are strongly hybridized near Fermi level, indicating strong bonding between Sn and Sb atoms in the KSnSb. The nature of bonding is mixed covalent and ionic type which can be understood by considering the K and SnSb layers individually. KSnSb primarily consists of a sp$^3$ type buckled honeycomb (similar to monolayer stanene) layer of [SnSb]$^{-\delta}$ charge configuration. [K]$^{+\delta}$ serve as stuffing layers and generate the net polarization along z-axis as shown in Fig.1(a) of the main manuscript. We first consider the stanene [Sn monolayer, See Fig.~\ref{charge}(a)], the band structure for which is shown in Fig.~\ref{charge}(c). The p-orbitals overlap of Sn1 and Sn2 (in a two atom unit cell) are isotropic i.e, the contribution of Sn1 p- and Sn2 p-orbitals are exactly same in the band structure. This indicates that the relative charges on Sn1 and Sn2 are same. Bader charge calculation also shows that the relative charges on both the Sn-sites are zero in stanene. Hence, the nature of chemical bonding in Sn monolayer is covalent. However, if we replace one Sn by one Sb atom, one extra electron comes in the SnSb system (see Fig.~\ref{charge}(b)). Indeed, Bader charge calculation shows that the relative charges of Sn and Sb are +0.5e and -0.5e which creates charge imbalance on these two atoms. The charge imbalance is clearly visible in the band structure of SnSb monolayer, as shown in Fig.~\ref{charge}(d), because the contribution of Sn and Sb p-orbitals near Fermi level (within -3 eV to 3 eV energy) are unequal. In case of KSnSb, K atom has cationic character with a deficit of one electron which is supposed to come from SnSb layer in case of strong ionic bonding between K and SnSb layers. However, Fig.~\ref{charge}(e) shows that overlap between K and SnSb layers are very small and the band gap is considerably smaller than a typical ionic crystals. This indicates that the bonding between K and SnSb layers are relatively weak. The overall charge configuration is K$^{+\delta}$[SnSb]$^{-\delta}$ where delta is smaller than 1. In case of KSnSb, Bader charge calculation shows that relative charge of K, Sn and Sb are +0.63e, +0.25e and -0.88e respectively and the value of ${\delta}$ is 0.63. 

\begin{figure}[b!]
\centering
\includegraphics[width=1\textwidth]{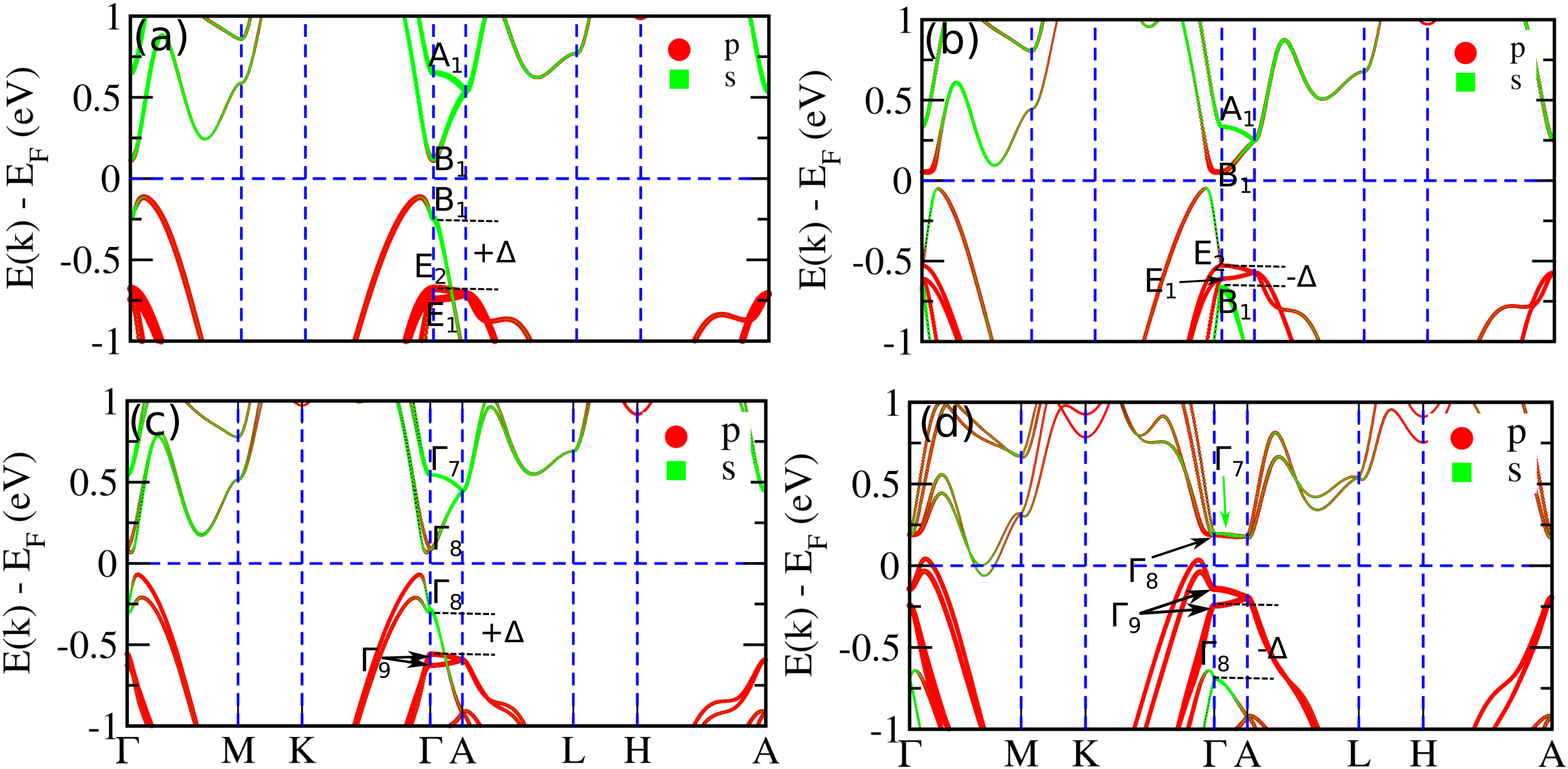}
\caption{(Color online) Band structures of KSnSb (left pannel) and KSnBi (right pannel) (a,b) without and (c,d) with SOC. Orbital projections and symmetry enforced band characters at $\Gamma$-point are denoted for all cases. For both compounds, A$_1$ transforms to $\Gamma_7$, B$_1$ transforms to $\Gamma_8$ and E$_1$, E$_2$ transform to $\Gamma_9$ band character when SOC is added.}
	\label{wosoc}
\end{figure}

\subsection{Band structures of KSnSb and KSnBi with and without SOC (Band Representation and Band Inversion Mechanism)}

\label{band-inversion}

In this section, we have compared the band structures of KSnSb and KSnBi with and without SOC to understand the band inversion mechanism properly. 
Figure \ref{wosoc}(a,b) show the band structures of KSnSb and KSnBi without (w/o) SOC. The corresponding band structures with SOC are shown in Fig.~\ref{wosoc}(c,d). The irreducible representation (IR) of the bands at $\Gamma$ point are indicated for both the compounds in both with and without SOC cases. A$_1$, B$_1$, E$_1$ and E$_2$ represents the IRs of C$_{6v}$ point group with single valued representation. Whereas $\Gamma_7$, $\Gamma_8$ and $\Gamma_9$ are the IRs when SOC is included (double valued representation). Table S2, S3 and S4 show the details of point group character tables. When SOC is included, the IRs transform as: A$_1\rightarrow$$\Gamma_7$, B$_1\rightarrow\Gamma_8$, E$_1\rightarrow\Gamma_9$, E$_2\rightarrow\Gamma_9.$  

In case of KSnSb, both CBM and VBM (at $\Gamma$ point) majorly consists of s-like orbital character in both w/o SOC (B$_1$ band representation) and with SOC ($\Gamma_8$ band representation) cases. p-orbital type E$_1$ and E$_2$ bands (below the B$_1$ VBM) transform to p-type $\Gamma_9$ bands when SOC is included. However, in case of KSnBi, the orbital characters (s- and p-orbitals) are inverted between CBM and VBM even without SOC due to crystal field splitting. Unlike KSnSb, the CBM of KSnBi possess p-type orbital contribution (B$_1$ band character). Including SOC, CBM B$_1$ bands transform to $\Gamma_8$ character under double group representation. Other B$_1$ band (below the E$_1$ and E$_2$ bands in valance band) of KSnBi retains its s-character which is sifted below the E$_1$ and E$_2$ bands due to crystal field splitting because of the larger size of Bi than Sb, as evident from Fig. \ref{wosoc}. Inclusion of SOC in KSnBi (Fig.~ \ref{wosoc}(d)) also increases the gap between $\Gamma_8$ and $\Gamma_9$ bands in valance band. So, in case of KSnSb, both the $\Gamma_8$ bands has s-like character while in case of KSnBi CBM $\Gamma_8$ band is p-type and another $\Gamma_8$ band  in valance band is s-type which confirms the band inversion.

The bands near the Fermi level (E$_F$) are governed by the s, p$_x$, p$_y$ and p$_z$ atomic orbitals of the Sn and Sb(Bi) atoms for both KSnSb and KSnBi. The representations of bands which take part in band inversion near E$_F$ are $\Gamma_8$ and $\Gamma_9$. The z-component of total angular momentum J$_z$ for these reps are $\pm$5/2 and $\pm$3/2 (for $\Gamma_8$ and $\Gamma_9$ bands). Corresponding symmetrized bases for these states can be chosen as; $\ket{j,m}$ =  $\ket{5/2, \pm 5/2}$ and $\ket{3/2, \pm 3/2}$.\cite{Altman1994} Using such symmetrized bases, low energy Hamiltonian of similar compound LiZnBi (186 space group) has been constructed in Ref.~[\onlinecite{LiZnBi}]. 

\subsection{HSE06 calculations of KSnSb and KSnBi}
\label{HSE06}

\begin{figure}[h!]
\centering
\includegraphics[width=0.6\textwidth]{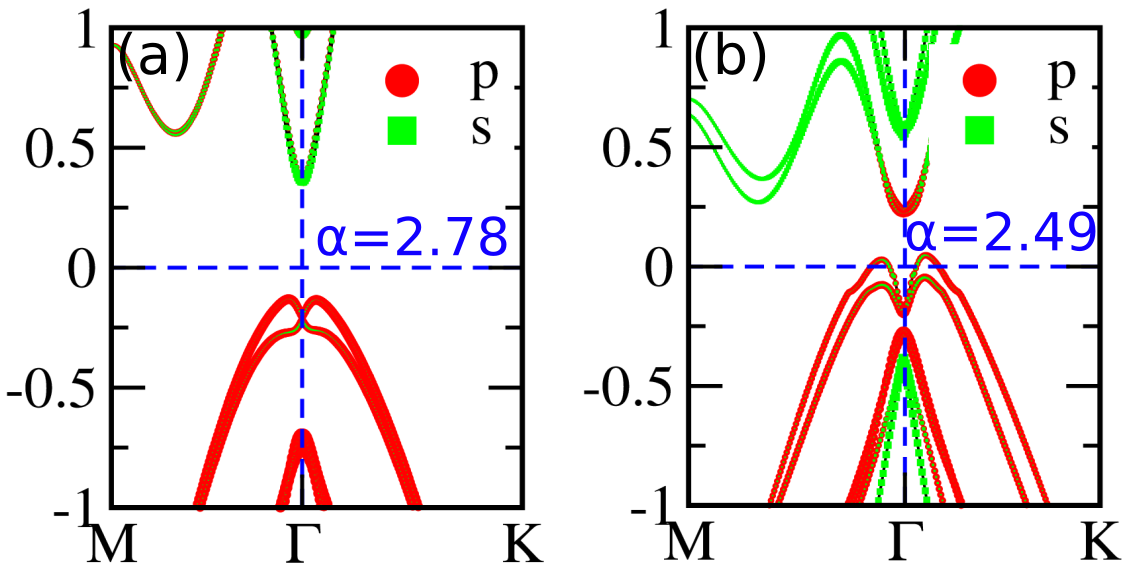}
\caption{(Color online) HSE06+SOC band structures of (e) KSnSb and (f) KSnBi along M-$\Gamma$-K direction.}
\label{hse06}
\end{figure}

GGA level of calculations often underestimate the band gap and therefor the true band ordering in a low gap material may shift from shift from the actual results. As such, for further verification of our predictions, we computed band structures of the parent compounds KSnSb and KSnBi using accurate HSE06 functional as shown in Fig.~\ref{hse06}(a,b). The HSE06 bands ordering are similar in both compounds with the GGA results.

\subsection{Topological index of KSnSb and KSnBi}
\label{Z2}

\begin{figure}[hb!]
	\centering
	\includegraphics[width=0.5\textwidth]{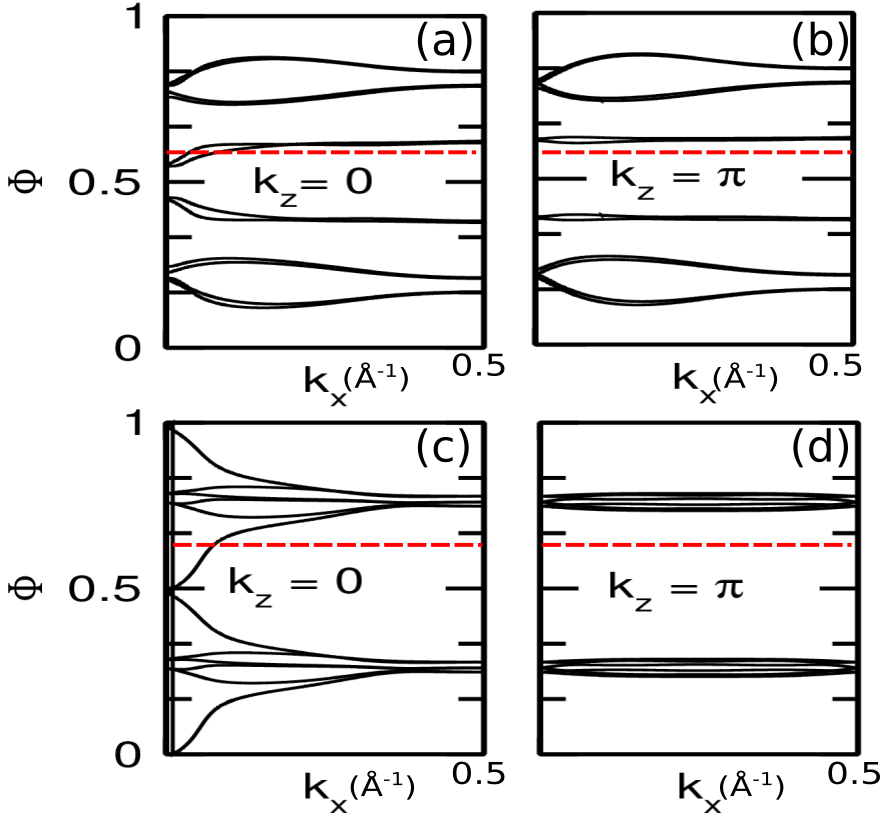}
	\caption{(Color online) Evolution of wannier charge center ($\Phi$) of (a,b) KSnSb and (c,d) KSnBi on $k_z$=0 and $k_z$=$\pi$ plane respectively. $\Phi$ is plotted in the units of \emph{z}-axis lattice constant \emph{c}. Here, in the 2D BZ, we plot $\Phi$ as a function of $k_x$ momenta and $k_y$ is integrated out while evaluating numerical value of WCC. }
	\label{z2}
\end{figure}


The topological $\mathbb{Z}_2$ invariant is calculated by observing the evolution of the so-called Wannier charge center (WCC) in a plane. Here we very briefly discuss about the WCC and it's evolution. A detailed pedagogical description about WCC and it's relation with the topology is already given in Refs.~\onlinecite{Z2WCC1,Z2WCC2,Z2WCC3}. For a periodic system, the electronic states are described by the Bloch function $|\psi_{n\mathbf{k}}\rangle = e^{i\mathbf{k\cdot r}}|u_{n\mathbf{k}}\rangle$, where $|u_{n\mathbf{k}}\rangle$ is the cell periodic part. Using Fourier transformation, one can define another alternative representation of Bloch functions which is nothing but a set of localized orbitals, called Wannier functions (WFs) and they are defined as--
\[|W_n\left( \mathbf{R}\right) \rangle =\frac{1}{(2\pi)^3}\int_{BZ}d\mathbf{k}~e^{i\mathbf{k\cdot (r-R)}} |u_{n\mathbf{k}}\rangle \]

Choosing a particular direction (say $\hat{z}$) of Wannierization in 3D, the WFs in a 2D plane takes the following form --
\begin{eqnarray}\label{2DWF}
|W_{n}\left( l_z,k_x,k_y\right)\rangle = \frac{1}{2\pi} \int dk_z~e^{i\mathbf{k}\cdot \left( \mathbf{r}-l_z c \hat{z}\right) } |u_{n\mathbf{k}}\rangle,
\end{eqnarray}
where $l_z$ is an integer representing the layer index and $c$ is the lattice constant along $\hat{z}$ direction. This particular WF is localized along $\hat{z}$, and therefore its position which is called WCC can be defined as --
\[
\bar{z}_n\left( k_x,k_y\right) = \langle W_{n}\left( 0,k_x,k_y\right)|\hat{z}|W_{n}\left( 0,k_x,k_y\right)\rangle 
\] 
The evolution of WCC shares the qualitative features of the surface dispersion $\epsilon_n(k_x,k_y)$,\cite{Z2WCC1,Z2WCC2,Z2WCC3} which is the manifestation of bulk-boundary correspondence. Tracking of WCC between two time reversal invariant momenta (TRIM) points in a 2D plane, could provide information about the trivial/non-trivial topology of the plane.\cite{Z2WCC1,Z2WCC2,Z2WCC3} Since, it has been shown that WCC and corresponding energy dispersion in a 2D plane are closely related to each other in terms of topology and symmetries of the system, therefore one can relate the WCC branches in Fig.~\ref{z2} with the edge states which we typically see in the trivial/non-trivial 2D insulating system.\cite{Z2WCC1} As in the case of 2D TI, where one counts the number of crossings by the edge states with the Fermi energy reference in order to understand the trivial/non-trivial topology of the system, here for WCC evolution, one can consider a generic reference line between $k_x=0$ to $k_x=\frac{\pi}{a}$ and can count the number of crossings by the WCC with the reference line. This counting of WCC crossings equivalently demonstrates the topological trivial or non-trivial structures.\cite{Z2WCC1,Z2WCC2,Z2WCC3}  Similar to the case of edge states, if the number of crossings is odd then the plane has non-trivial topology, i.e., $\mathbb{Z}_2=1$, otherwise $\mathbb{Z}_2=0$.  \cite{hasan2010}

Figure~\ref{z2} shows the evolution of WCC ($\Phi$) along $k_x$ axis in $k_z$=0 and $k_z$=$\pi$ plane for both the compounds KSnSb and KSnBi. As described above as well as in Refs.[\onlinecite{Z2WCC1,Z2WCC2,Z2WCC3}], it is obvious that KSnSb is a trivial insulator as there are no branches of WCC which encounter any odd number of jump across any generic reference line in both $k_z$=0 and $k_z$=$\pi$ plane. However, KSnBi is topologically non-trivial ($\mathbb{Z}_2= 1$), because the WCC at least in $k_z=0$ plane, cuts a generic reference line (dashed horizontal line) odd number of times in the BZ, as shown in Fig.~\ref{z2}(c).

\subsection{Band gap evolution under pressure for KSnSb and KSnBi}
\label{band-gap-evoluation}

\begin{figure}[hb!]
\centering
\includegraphics[width=0.5\textwidth]{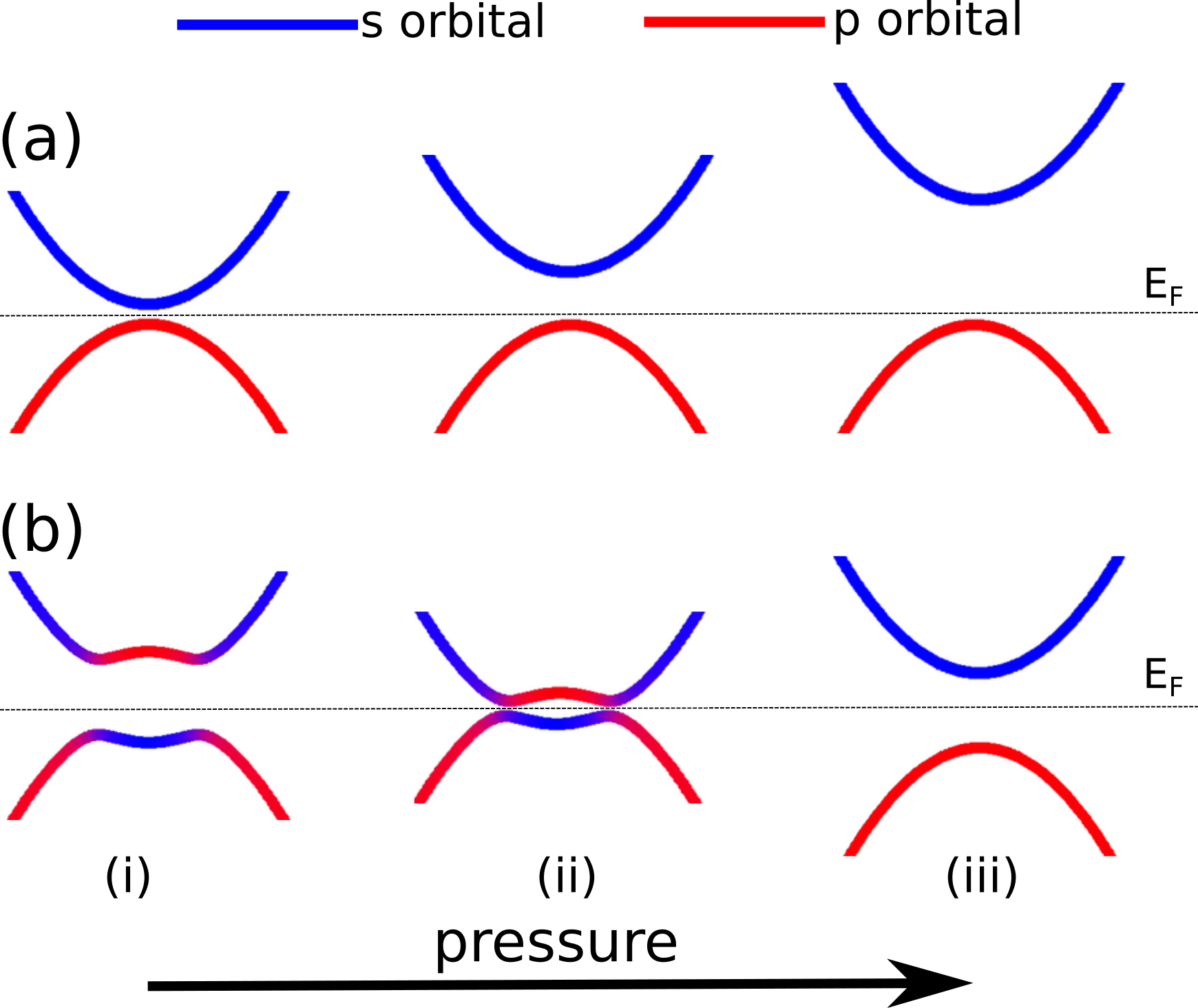}
\caption{(Color online) (a) (top panel) Evolution of band gap under pressure for a trivial insulator (KSnSb) and (b) (bottom panel) for a topological insulator (KSnBi).}
\label{ti-gap}
\end{figure}

Lets now discuss about the different behavior of band gap for the two parent systems under pressure. It is to be noted that KSnSb is a trivial insulator while KSnBi is a topological insulator (TI). Let us first discuss, in general, how differently the band gap changes for a TI as compared to a trivial insulator under compressive pressure. Figure ~\ref{ti-gap} shows a schematic representation of the evolution of bands under pressure for a trivial insulator (Fig. ~\ref{ti-gap}(a)) and a TI (Fig. ~\ref{ti-gap}(b)). In both the cases, s-orbital gets destabilized under pressure and shift towards the higher energy. For a trivial insulator (such as KSnSb), p-orbitals mediated bands (lying below E$_F$) hardly get affected and hence the increase in band gap is purely controlled by the upward shift of s-band. For a TI (such as KSnBi) however, the s-p bands are already inverted at ambient condition. Under pressure, s-orbital shifts upward and p-orbital slowly comes downward in energy such that the two bands meet at a critical pressure (Fig. ~\ref{ti-gap}(b-ii)) and then reopen (Fig. ~\ref{ti-gap}(b-iii)) with further increase in pressure. This is a well know adiabatic transformation for any TI, which occur when the compound undergoes a topological phase transition from non-trivial to trivial phase.

\begin{figure}[h!]
\centering
\includegraphics[width=0.8\textwidth]{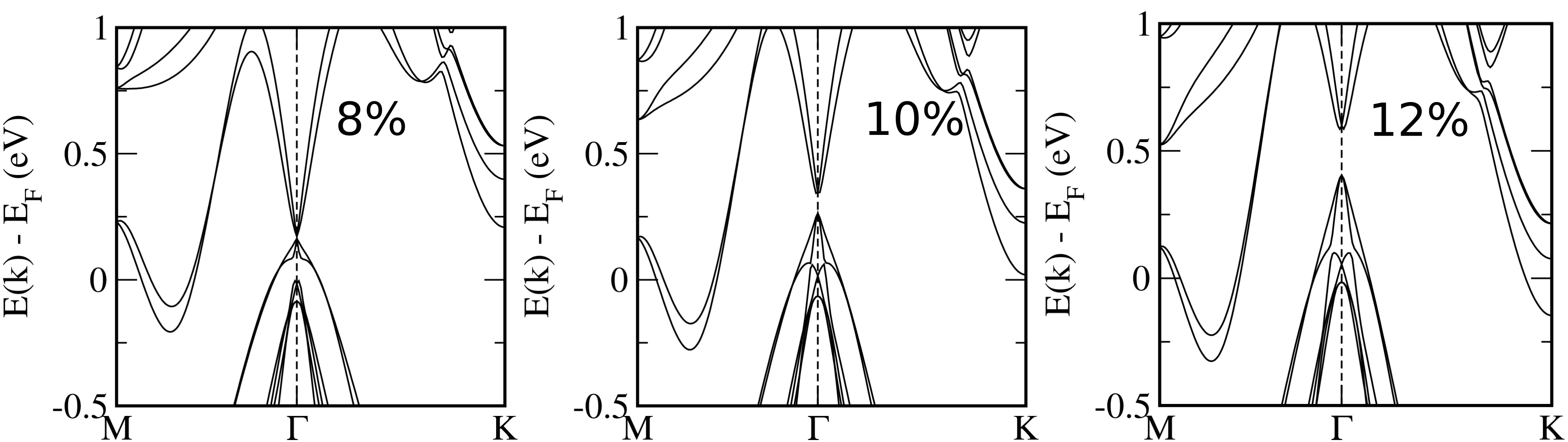}
\caption{(Color online) Band structures of KSnBi under 8\%, 10\% and 12\% lattice strain}
\label{KSnBi-high-pressure}
\end{figure}

Now, coming back to the real system KSnSb, the symmetry character of both the conduction band (CB) and valance band (VB) near $\Gamma$-point are $\Gamma_8$-type. But in terms of orbital character, $\Gamma_8$ CB is predominantly contributed by s-like orbitals of Sn and Sb atoms whereas $\Gamma_8$ VB is contributed by p-like orbitals of Sn and Sb (see Fig. 1(c) of main manuscript). In KSnSb, the K$^{2+}$ cations are staffed in between [SnSb]$^{2-}$ buckled honeycomb network (see Fig. 1(a) in main manuscript). As such, s-orbitals of Sn \& Sb lie outside the p orbitals. With increasing pressure, s orbitals get destabilized and shift towards the higher energy. However, the p-like CB remains almost unaffected. As such, the band gap increases trivially under compressive pressure. The scenario for KSnBi is similar to the bottom panel, Fig. ~\ref{ti-gap}(b)). It closes its topological gap at 6\% lattice strain (see Fig. 3(f) of the main manuscript) and then reopen a trivial gap at higher pressure at $\Gamma$ point. Figure ~\ref{KSnBi-high-pressure} shows the band structures of KSnBi at further larger lattice strain of 8\%, 10\% and 12\%.

\subsection{In-plane spin texture of KSnSb$_{0.5}$Bi$_{0.5}$}
\label{in-plane}

\begin{figure}[h!]
\centering
\includegraphics[width=0.75\textwidth]{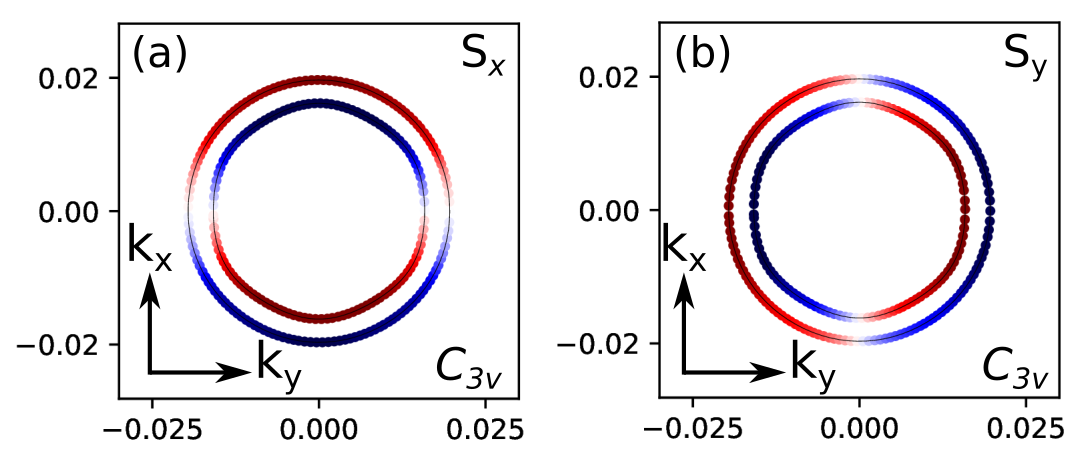}
\caption{(Color online) In-plane S$_x$ and S$_y$ components of KSnSb$_{0.5}$Bi$_{0.5}$ at 250 meV energy cut.}
\label{sxsy}
\end{figure}

In this section we have shown the in-plane spin texture i.e, S$_x$ and S$_y$ components of KSnSb$_{0.5}$Bi$_{0.5}$. Figure~\ref{sxsy}(a,b) shows the S$_x$ and S$_y$ components of KSnSb$_{0.5}$Bi$_{0.5}$ respectively. 

\begin{figure}[h!]
	\centering
	\includegraphics[width=0.4\textwidth]{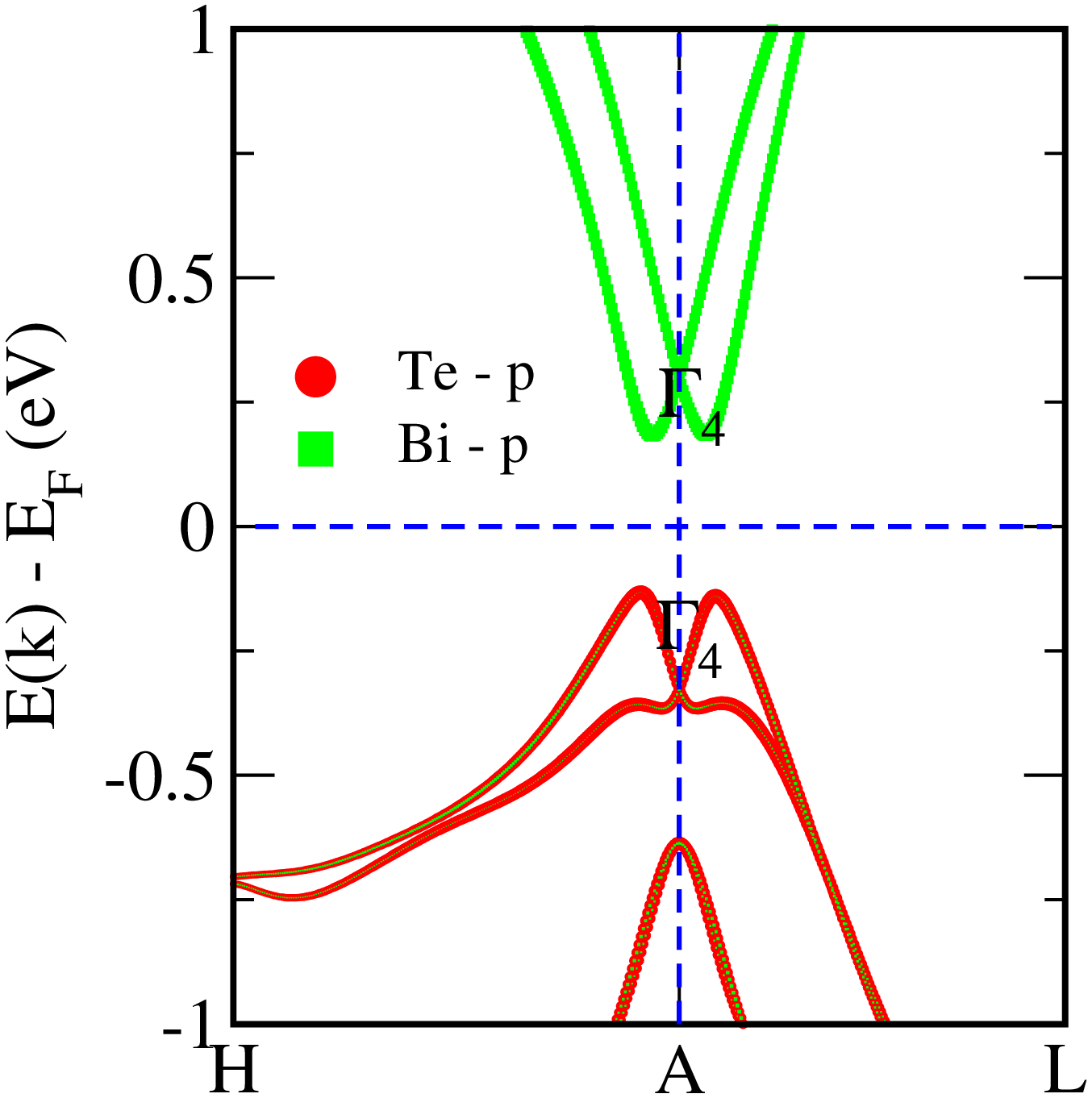}
	\caption{(Color online) Band structure of BiTeI calculated within GGA approximation. $\Gamma_4$ denotes the band symmetries for conduction and valance bands.}
	\label{BiTeI}
\end{figure}

\subsection{Band structure of BiTeI}
\label{org}
Within GGA approximation, we have calculated the band structure of the extensively studied Rashba semiconductor BiTeI to validate our proposed criteria for getting large Rashba parameters (Rashba splitting $\Delta E$ and $\alpha$) in this compound. Figure~\ref{BiTeI} shows the band structure of BiTeI along H-A-L line on $k_z$ = $\pi$ plane. It is clear from the figure that both the conduction and valence band possess similar band characters at A point. As such the term $<u_n|\frac{\hbar}{4 m_0^2 c^2}[{\bf \nabla{V}\times p}]\cdot\sigma|u_m>$ is symmetrically allowed which in turn contribute to the Rashba splitting energy $\Delta E$ and provide giant Rashba coefficient $\alpha$.

\begin{table}[h]
\begin{ruledtabular}
\caption{Rashba parameters of conduction band for different materials; Rashba energy ($\Delta E$), momentum offset ($\Delta k$) and Rashba coefficient ($\alpha$). Trivial insulator (NI), topological insulator (TI) and Weyl semimetal (WSM) represent the various topological phases (T-phase) of compounds.}
\label{Table1}
\begin{tabular}{c c c c c c c c c}
& Compound &   $\Delta E$ (meV)  & $\Delta k$ ($\AA$)      & $\alpha$ (eV$\AA$) & T-phase              \\ \hline
& KSnSb  $^a$                                            &    19    &  0.028  &      1.29         & NI  \\
& KSnSb$_{0.875}$Bi$_{0.125}$ $^a$                       &    23    &  0.035  &      1.34         & NI  \\
& KSnSb$_{0.75}$Bi$_{0.25}$ $^a$                         &    32    &  0.047  &      1.37         & NI   \\
& KSnSb$_{0.625}$Bi$_{0.375}$ $^a$                       &    41    &  0.062  &      1.33         & WSM   \\
& KSnSb$_{0.5}$Bi$_{0.5}$  $^a$                          &    23    &  0.052  &      0.87         & TI    \\
& KSnBi  $^a$                                            &    7     &  0.120  &      1.17         & TI    \\ 
\end{tabular}
\end{ruledtabular}
$^a$ Reference [This Work].
\end{table}

\subsection{Character Table for C$_{6v}$ and C$_s$ point groups}
\label{CT}

\begin{table}[h]
	\begin{ruledtabular}
		\caption{C$_{6v}$ point group character table}
		\label{Table3}
		\begin{tabular}{c c c c c c c c c c c c}
			& C$_{6v}$ &  E  &   2C$_6$  & 2C$_3$ & C$_2$ &  3$\sigma_v$ & 3$\sigma_d$ &   \\ \hline	
			
			& A$_1$ &  1  & 1 & 1 & 1 & 1 & 1 &   \\
			
			& A$_2$ &  1  & 1 & 1 & 1 & $-$1 & $-$1 &   \\
			
			& B$_1$ &  1  & $-$1 & 1 & $-$1 & 1 & $-$1 &   \\
			
			& B$_2$ &  1  & $-$1 & 1 & $-$1 & $-$1 & 1 &     \\
			
			& E$_1$ &  2  & 1 & $-$1 & $-$2 & 0 & 0  &      \\
			
			& E$_2$ &  2  & $-$1 & $-$1 & 2 & 0 & 0  &      \\

		\end{tabular}
	\end{ruledtabular}
\end{table}

\begin{table}[h]
	\begin{ruledtabular}
		\caption{C$_{6v}$ double group character table}
		\label{Table2}
		\begin{tabular}{c c c c c c c c c c c c}
			& C$_{6v}$ &  E  & $-$E &  2C$_6$ &  $-$2C$_6$  & 2C$_3$ & $-$2C$_3$  & C$_2$ &  3$\sigma_v$ \& $-$3$\sigma_v$ & 3$\sigma_d$ \& -3$\sigma_d$&   \\ \hline	
			
			& $\Gamma_7$ &  2  & $-$2 & $-\sqrt{3}$ &  $\sqrt{3}$ & 1 & $-$1  & 0 &  0 & 0 &  \\	
			
			& $\Gamma_8$ &  2  & $-$2 & $\sqrt{3}$ &  $-\sqrt{3}$ & 1 & $-$1  & 0 &  0 & 0 &  \\
			
			& $\Gamma_9$ &  2  & $-$2 & 0 &  0 & $-$2 & 2  & 0 &  0 & 0 &  \\	
		\end{tabular}
	\end{ruledtabular}
\end{table}

\begin{table}[hb!]
	\begin{ruledtabular}
		\caption{C$_s$ double group character table}
		\label{Table2}
		\begin{tabular}{c c c c c c c }
			& C$_s$ &  E  & $-$E &  $\sigma_v$ &  $-\sigma_v$ &   \\ \hline	
			
			& $\Gamma_1$ &  1  & $-$1 & i &  $-$i & \\	
			
		    & $\Gamma_2$ &  1  & $-$1 & $-$i &  i & \\
			
		\end{tabular}
	\end{ruledtabular}
\end{table}


 
